\documentclass[]{Liebert_Author}

\title{Fairness Evaluation for Uplift Modeling \\ in the Absence of Ground Truth
} 

\author{
Serdar Kad{\i}o\u{g}lu,$^{1, 2,\ast}$, 
Filip Michalsk\'y$^{1}$ \\
{$^{1}$AI Center of Excellence, Fidelity Investments}\\
{245 Summer St, Boston MA 02318, USA}\\
{$^{2}$Dept. of Computer Science, Brown University}\\
{Providence, RI 02912, USA}\\
{$^\ast$To whom correspondence should be addressed;}\\
{E-mail: serdar.kadioglu@fmr.com}
}

\usepackage{xcolor}
\usepackage{booktabs}
\usepackage{multirow}
\usepackage{rotating,tabularx}
\usepackage{amsthm}
\theoremstyle{definition}
\newtheorem{definition}{Definition}[section]

\begin{document} 

\maketitle 


\begin{abstract}

The acceleration in the adoption of AI-based automated decision-making systems poses a challenge for evaluating the fairness of algorithmic decisions, especially in the absence of ground truth. When designing interventions, uplift modeling is used extensively to identify candidates that are likely to benefit from treatment. However, these models remain particularly susceptible to fairness evaluation due to the lack of ground truth on the outcome measure since a candidate cannot be in both treatment and control simultaneously. In this article, we propose a framework that overcomes the missing ground truth problem by generating surrogates to serve as a proxy for counterfactual labels of uplift modeling campaigns. We then leverage the surrogate ground truth to conduct a more comprehensive binary fairness evaluation. We show how to apply the approach in a comprehensive study from a real-world marketing campaign for promotional offers and demonstrate its enhancement for fairness evaluation.
\end{abstract}

\section{Introduction}

Uplift modeling helps decide interventions that would drive the most positive outcomes. Prominent applications of uplift modeling include Personalized Healthcare, Advertising, and Marketing~\cite{BoothDavid2019Mait, MariAlex2019}. In these domains, it is of great practical interest to identify users that exhibit certain positive behaviors and are more likely to respond and benefit from an intervention. As an example, the positive outcome can follow business performance metrics such as purchasing a product or signing up for an offer, which can be boosted by an intervention. Notice that some users might already be inclined to those outcomes naturally \textit{without} an intervention, and conversely, there might be some other users who would be adversely impacted by the intervention causing them not to take action. The main idea behind \textit{campaign design} is to find users who would truly benefit from the intervention. 

\smallskip

\noindent\rule{345pt}{0.5pt}

\noindent {This article is an extended version of the short paper published in the proceedings of ICMLA 2021, “Surrogate Ground Truth Generation to Enhance Binary Fairness Evaluation in Uplift Modeling"~\cite{9680169}.}

\noindent\rule{345pt}{0.5pt}

\smallskip

Notably, \textit{uplift modeling} has been a popular choice in designing successful campaigns. Uplift models are employed extensively in e-commerce, personalized medicine, and customer relationship management for up-sell, cross-sell, and retention management, see, e.g., ~\cite{gubela2019,KaneKathleen2014Mftt} for a comprehensive review of uplift modeling and their applications.

In parallel, it is well-documented that predictive models suffer from historical bias in training data concerning protected attributes~\cite{propublica, debias, pmlr-v81-buolamwini18a}. These attributes can be race, gender, age, or marital status that are legally protected in domains such as housing~\cite{fairhousing}, credit systems~\cite{fairlending}, and human resources~\cite{fairhiring}. Recent studies revealed algorithmic bias in several applications, including healthcare~\cite{optumbias}, credit offers~\cite{loan-ai}, university admissions~\cite{barocas-hardt-narayanan} and hiring decisions~\cite{hbs_hiring_algos}.


To mitigate bias when present, several methods have been proposed over the recent years from pre-processing the  data~\cite{BolukbasiTolga2016QaRS,CalmonFlavioP.2017ODPf}, to in-processing the model~\cite{barocas2016big, dwork2012fairness, equal_of_op}, and post-processing the predictions~\cite{pmlr-v81-menon18a}. 
Independent of the intervention phase, our ability to evaluate algorithmic fairness remains key toward trusted decision making. Several fairness metrics are devised to assess algorithmic bias~\cite{beretta2019, Corbett-DaviesSam2017Adma} and to avoid unwanted consequences. 
In the context of campaign design, \textit{binary} fairness metrics are immediately relevant for decisions of the form contact/no-contact, offer/no-offer.  The important realization is that most binary fairness metrics require \textit{ground truth data}. This poses a significant bottleneck for fairness evaluation when true labels remain unobservable. Consequently, model predictions cannot be evaluated with respect to fairness, which increases the risk of deploying AI applications that are not tested rigorously.


Campaign design for intervention is particularly prone to the lack of such ground truth. At a high-level, uplift models are trained based on historical responses from \textit{treatment} and \textit{control} groups. Then, the model is used to differentiate between intervention vs. no-intervention. In practice, a campaign can be deemed successful when it performs better than the alternative approach or the business-as-usual toward a desired outcome. However, once the campaign period is over, it remains unknown, even when the campaign is successful, whether the intervention/no-intervention decision was the best possible action for every individual. That is, there exists no ground truth that defines the \textit{best} possible campaign design. Critically,  in the absence of such ground truth, our ability to conduct algorithmic fairness evaluation is severely limited since most binary fairness metrics require true labels. This is exactly the problem we address. We extend our framework proposed in~\cite{9680169} with additional details on the algorithm and background, and more comprehensive evaluations. The framework allows generating surrogate ground truth for uplift models used in campaign design. In turn, the surrogate ground truth enables a more extensive binary fairness evaluation as demonstrated on a real-world campaign for promotional offers. 

\section{Problem Definition}

Let us start with a description of our setting that consists of the Campaign Design Problem:

\theoremstyle{definition}
\begin{definition}[\textbf{Campaign Design Problem (CDP)}]\label{problem}
Given a population $P$, we are interested in running a campaign, $M$, to target a subset of the population between the time period $t_{\text{start}}$ and $t_{\text{end}}$ to drive a key performance indicator (KPI) $K$. 
\end{definition}

The goal of the Campaign Design Problem (CDP) is to find the most suitable subset of the population $P$ for an intervention using an uplift modeling approach based on treatment and control groups, $T$, and $C$, respectively. Optionally, the campaign can be subject to certain capacity or budgetary restrictions, denoted by $B$, that might constrain the number of interventions.

Notice how the $CDP \langle P, T, C, M, K, B \rangle$ captures practical applications in various domains. For example, in marketing, we might be interested in identifying users for a promotional product offering, or in advertising, we might be interested in users who would benefit the most from our recommendations. Using an uplift model that estimates the effect of treatment vs. control, at time $t_{start}$, we enroll users in the campaign with the highest lift (i.e., treatment minus control effect) without violating the budget constraint $B$ (e.g., a limit on the number of free products to give away). Then, at time $t_{end}$, we measure the campaign's overall success for the KPI $K$. However, while it is possible to evaluate whether the overall campaign was successful, we do not know whether intervention vs. no-intervention was the best decision for each individual. Notice how we cannot simultaneously enroll and not enroll a user in the campaign and cannot directly measure what would happen if the user had been in the complementary group. As such, the \textit{best possible} campaign design remains unknown, and there is no ground truth. 

\section{Our Contribution}
Most binary fairness metrics require truth labels, as discussed in more detail in next sections. This dependency on truth labels makes the case of CDP highly problematic from an AI Bias perspective. Any application of the CDP in practice suffers from the absence of ground truth, and hence, lacks a holistic AI fairness evaluation. To address this challenge, the main contribution of this paper is to propose an algorithm for Surrogate Ground Truth (SGT) generation for the CDP. Our approach is generic and  does not make any assumptions on the training data or the uplift modeling approach used. The immediate benefit of the surrogate ground truth is enhanced binary fairness evaluations which would not be possible otherwise. 

In the remainder of this paper, we start with a brief background on uplift modeling (\textsection Section~\ref{uplift}) to solve the CDP. We then provide an overview of commonly used binary fairness metrics (\textsection Section~\ref{metrics}) and highlight their dependency on truth labels. We then present the details of our algorithm to generate surrogate ground truth (\textsection Section~\ref{sgt}). Using a publicly available marketing campaign dataset from Starbucks Promotional Offers, we demonstrate how the approach can be applied in practice (\textsection Section~\ref{Experimental}). Let us stress that the goal of our experiments is neither to build the best uplift model for this dataset nor highlight bias. Our main contribution is an algorithmic approach for generating surrogate ground truth, and our experiments demonstrate how to bring its components together to utilize the method in practice. We hope that researchers and data practitioners can map their scenarios to this setting and benefit from enhanced fairness evaluations in their applications.

\section{Background}\label{background}

\subsection{Uplift Modeling}\label{uplift}

Let us start with some basics on uplift modeling. For a more in-depth overview, we refer to~\cite{KaneKathleen2014Mftt, gubela2019}. The goal of uplift modeling is to predict the \emph{incremental impact} of a decision for each individual toward an outcome of interest. Each individual falls into one of the four segments with respect to an intervention~\cite{gubela2019}:

\begin{itemize}
    \item \textbf{Sure Things}: Positive action, such as a product purchase, is taken regardless of intervention.
    \item \textbf{Lost Causes}:  No positive action taken regardless of intervention.
    \item \textbf{Do-not-Disturbs}: Positive action taken unless disturbed.
    \item \textbf{Persuadables}: Positive action is taken only with intervention. 
\end{itemize}

\begin{figure}
\centering{
\includegraphics[height=1.9in,width=2.5in,keepaspectratio]{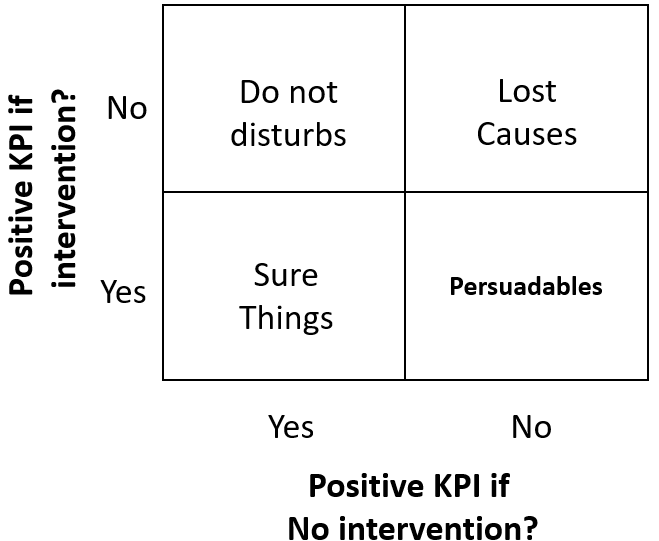} 
\caption{Four types of individuals in campaigns segmented by their response to an intervention. Adapted from~\cite{lo2020}.}}
\label{fig:one}
\end{figure}

The target segment is the \textit{persuadables}, who would not take action without the intervention. Uplift modeling helps in finding the persuadable group. There are three main uplift modeling approaches~\cite{pachamova2020}:

Figure~\ref{fig:one} captures these four quadrants of individuals with respect to positive KPI with and without intervention. Next, we cover a number of techniques that are commonly used in the literature for uplift modeling. 

\subsubsection{Two-model approach}
In this approach, we divide the population of interest into a treatment group $T$ and a control group $C$. Then, we train a model $M_T$ on group $T$ using features $X$ to estimate the probability of response if given treatment $\hat{p}(R|X, T)$. Similarly, we train a model $M_C$ on group $C$ using features $X'$ to estimate the probability of response without treatment $\hat{p}(R|X',C)$. Note that $X, X'$ may or may not overlap. Then, for an individual $i$, we estimate the incremental lift in probability to respond to an intervention $\hat{\pi}_i$ as:

\begin{equation}
    \hat{\pi}_i = \hat{p}(R|X, T) - \hat{p}(R|X',C)
\end{equation}

\subsubsection{Dummy-variable approach}
In this approach, we use a single model but include a dummy binary variable $D$ that denotes whether an individual was in the treatment or the control group. Then, the estimate of the incremental lift in probability to respond to an intervention $\hat{\pi}_i$ is calculated as: 

\begin{equation}
    \hat{\pi}_i = \hat{p}(R|X, D=1) - \hat{p}(R|X,D=0)
\end{equation}

\subsubsection{Four-quadrant approach}
The idea behind this approach is to turn the problem into a four-class classification task with labels \{Sure Things, Lost Causes, Do-not-Disturbs, Persuadables\}. The goal is then to predict the class of a given individual and focus on the persuadable group. 

\subsection{Binary Fairness Metrics}\label{metrics}

The CDP inherently involves a binary decision about the intervention, hence we focus on binary fairness metrics. There exist various binary fairness metrics in the literature, see, e.g. \cite{Corbett-DaviesSam2017Adma, FeldmanMichael2014Card, HajianSara2011Dpid, HajianSara2016ABFD, equal_of_op, JosephMatthew2016FiLC, speicher2018a} with no consensus on a single definition that is universally accepted. In fact, the impossibility theorem proves that some metrics are incompatible with each other and cannot be satisfied simultaneously~\cite{KleinbergMR17}. Here, without loss of generality, we consider a set of six commonly used binary fairness metrics. These metrics are readily available in fairness evaluation libraries such as Jurity: Recommenders \& Fairness Evaluation library~\cite{Jurity}. Among those, only Statistical Parity (SP) and Disparate Impact (DI) can be calculated without truth labels. All other metrics require ground truth for evaluation. That is, a campaign cannot be evaluated from a fairness perspective using any of these metrics.

\smallskip

We follow the definition of group fairness from~\cite{Corbett-DaviesSam2017Adma}. Consider a set of independent features $X \in \mathcal{R}$ and a dependent measure $Y \in [0,1]$. Let $A \in [0,1]$ denote the membership in a group of interest. In practice, the group membership often refers to a protected attribute.

\smallskip 
 
\noindent \textbf{Average Odds (AO):} 
 The average odds denotes the average of difference in false positive rate (FPR) and true positive rate (TPR) for two groups:
    \begin{equation}
    \frac{1}{2}[(FPR_{\text{group 1}}-FPR_{\text{group 2}})+(TPR_{\text{group 1}}-TPR_{\text{group 2}})]
    \end{equation}

\medskip

\noindent \textbf{Disparate Impact (DI)} measures the ratio of probabilities for predicted positive outcomes among the two membership groups.
    \begin{equation}
    \label{metric_di}
    \frac{P(\hat{Y}=1 | A = 0)}{P(\hat{Y}=1 | A = 1)}
    \end{equation}

\medskip

\noindent \textbf{Equal Opportunity (EO)} measures the equality (or difference) of True Positive Rate (TPR) stratified by group membership.
    \begin{equation}
    TPR_{\text{group 1}} - TPR_{\text{group 2}}
    \end{equation}
    
\noindent \textbf{FNR Difference (FNR Diff)} measures the difference in False Negative Rates stratified by group membership.
    \begin{equation}
    FNR_{\text{group 1}} - FNR_{\text{group 2}}
    \end{equation}
    
\medskip

\noindent \textbf{Predictive Equality (PE)} measures the difference in False Positive Rates stratified by group membership. This metric is symmetrical to $1$ - FNR Difference.
    \begin{equation}
    FPR_{\text{group 1}} - FPR_{\text{group 2}}
    \end{equation}

\medskip

\noindent \textbf{Statistical Parity (SP)} measures the difference of probabilities for a predicted positive outcomes among the two membership groups.
    \begin{equation}
    P(\hat{Y}=1 | A = 0) - P(\hat{Y}=1 | A = 1)
    \end{equation}

Table~\ref{metrics-summary} lists the binary fairness metrics, ideal values, and their dependency on truth labels. While there could be different interpretations of the fairness values there is one guideline  established jointly by the U.S. Civil Service Commission, the Department of Labor, the Department of Justice, and the Equal Employment Opportunity Commission. This is captured as part of the Uniform Guidelines on Employee Selection Procedures~\cite{ug2020}, which defines a ``substantially different decision'' as the ``80\% rule'', where it is desirable to have decisions that are at least 80\% similar across groups. For example, if the ideal value of a metric is $0$, we consider an outcome in accordance with the metric when it falls between $-0.2$ and $0.2$. In Table~\ref{metrics-summary}, this guideline is captured by the lower and upper bound columns for every binary metric. 

\begin{table}[t]
\centering{
\renewcommand{\arraystretch}{1.3}
\caption{Binary fairness metrics and their dependency on truth labels.}\label{metrics-summary}
\begin{tabular}{ccccc}
\hline
\multicolumn{1}{c}{\begin{tabular}[c]{@{}c@{}}Binary \\ Fairness Metric\end{tabular}}              & \multicolumn{1}{c}{\begin{tabular}[c]{@{}c@{}}Ideal \\ Value\end{tabular}} & \multicolumn{1}{c}{\begin{tabular}[c]{@{}c@{}}Lower\\ Bound\end{tabular}} & \begin{tabular}[c]{@{}c@{}}Upper\\ Bound\end{tabular} & \begin{tabular}[c]{@{}c@{}}Requires \\ Label\end{tabular} \\ \hline
Average Odds        & 0                                                                          & -0.2                                                                      & 0.2                                                   & Yes                                                    \\
Disparate Impact    & 1                                                                          & 0.8                                                                       & 1.2                                                   & No                                                     \\
Equal Opportunity   & 0                                                                          & -0.2                                                                      & 0.2                                                   & Yes                                                    \\
FNR difference      & 0                                                                          & -0.2                                                                      & 0.2                                                   & Yes                                                    \\
Predictive Equality & 0                                                                          & -0.2                                                                      & 0.2                                                   & Yes                                                    \\
Statistical Parity  & 0                                                                          & -0.2                                                                      & 0.2                                                   & No                                                     \\ \hline
\end{tabular}
}
\end{table}

\begin{figure*}[t]
	\begin{center}
\includegraphics[width=145mm,height=120mm,keepaspectratio]{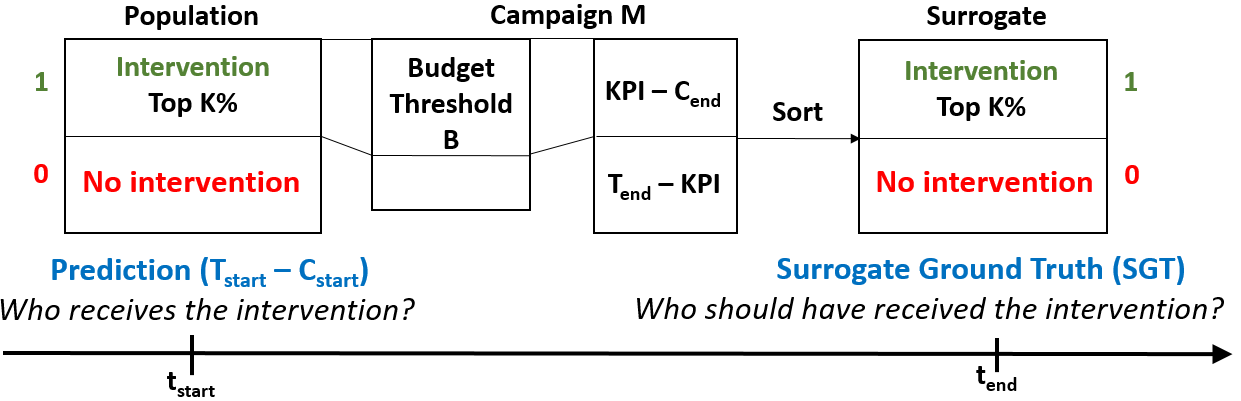}
\caption{The illustration of the Surrogate Ground Truth (SGT) generation algorithm.}\label{sgt-diagram}
	\end{center}
\end{figure*}

\section{Surrogate Ground Truth}\label{sgt}

Given an instance of the $CDP \langle P, T, C, M, K, B \rangle$, we generate the Surrogate Ground Truth (SGT) as illustrated in Figure~\ref{sgt-diagram} and detailed in Algorithm~\ref{algo}.

At a high-level, our algorithm consists of two main steps. In the first step, at time $t_{start}$, we identify a sub-population to receive treatment using the uplift strategy with treatment and control models $T$ and $C$, respectively. In the second step, at time $t_{end}$ and based on the campaign $M$ results, we perform a re-scoring operation to obtain the surrogate ground truth. 

\begin{algorithm}[p]
\SetAlgoLined
\SetKwInOut{Input}{input}\SetKwInOut{Output}{output}
\Input{Population $P$\\ 
Uplift models $T$ and $C$ trained to make inference on treatment and control groups \\
Campaign $M$ with KPI $K$ for experimentation between the time period $t_{start}$ and $t_{end}$ \\
Resource constraint $B$ on the budget of the intervention (optional)
}
\Output{Surrogate Ground Truth (SGT) for population $P$}

\texttt{\\}
// \textbf{Step - I}\\
\textbf{At time $t_{start}$}\;
$Lift_{start} \leftarrow \emptyset$\;
\For{user $\in$ population $P$}{
  $T_{start} \leftarrow$ T(user)\;
  $C_{start} \leftarrow$ C(user)\;
  $lift_{user} \leftarrow T_{start} - C_{start}$\;
  $Lift_{start} \leftarrow Lift_{start} \cup lift_{user}$\;
}

\texttt{\\}
$Ranking_{start} \leftarrow \text{sort\_descending}(Lift_{start})$\;
$size \leftarrow \text{select subset population size according to}\: B$\;
$population_{intervention} \leftarrow \text{users} \in Ranking_{start}[:size]$\;
$population_{no\_intervention} \leftarrow \text{users} \in Ranking_{start}[size:]$\;
\textbf{Run} campaign $M$ on $population_{intervention}$\;

\texttt{\\}
// \textbf{Step - II}\\
\textbf{At time $t_{end}$}\;
$Lift_{end} \leftarrow \emptyset$\;
\For{user $\in$ Population $P$}{
 $KPI_{user} \leftarrow query(KPI[user])$\;
  \eIf{user $\in$ \emph{$population_{intervention}$}}{
   $C_{end} \leftarrow C(user)$\;
   $surrogate_{user} \leftarrow KPI_{user}-C_{end}$\;
   }{
   $T_{end} \leftarrow T(user)$\;
   $surrogate_{user} \leftarrow T_{end}-KPI_{user}$\;
  }
  $Lift_{end} \leftarrow Lift_{end} \cup surrogate_{user}$\;
}

\texttt{\\}
$Ranking_{end} \leftarrow \text{sort\_descending}(Lift_{end})$\;
$surrogate_{intervention} \leftarrow \text{users} \in Ranking_{end}[:size]$\;
$surrogate_{no\_intervention} \leftarrow \text{users} \in Ranking_{end}[size:]$\;
$SGT \leftarrow surrogate_{intervention} \cup surrogate_{no\_intervention}$\;

\texttt{\\}
\Return $SGT$

\caption{Surrogate Ground Truth Generation}
\label{algo}
\end{algorithm}

\vspace{-0.2cm}
\subsubsection*{Step - I: Beginning of the Campaign}
\vspace{-0.2cm}
 At the beginning of the campaign, $t_{start}$, we use the predictions from the uplift model to decide who in the population should receive the intervention. Given the lift scores for the population, the $Top \; K\%$ can be selected for the intervention, e.g., to receive a promotional offer. 
In practice, the intervention decision can further be subject to resource constraints or budgetary considerations, e.g., a limit on the number of available promotions. A combination of uplift predictions, $Top \; K\%$ decile cutoff, and business constraints yields the selection for the treatment, and the campaign $M$ goes into effect. 

\subsubsection*{Step - II: End of the Campaign}
At the end of the campaign, $t_{end}$, we observe the results for the KPI $K$. At the heart of the SGT algorithm is a re-scoring operation as follows: 

\begin{itemize}
    \item The individuals from the original treatment group at time $t_{start}$ are re-evaluated according to the control model $C$ based on their features at time $t_{end}$. The surrogate lift of a treated individual is the difference between the observed KPI and the prediction from the \textit{control model}.
    \item  Analogously, the individuals from the original control group are re-evaluated using the treatment model $T$ based on their features at time $t_{end}$. The surrogate lift of a control individual is the difference between the prediction from the \textit{treatment model} and the observed KPI. 
\end{itemize}

Once the re-scoring operation generates surrogate lift for the entire population, individuals are re-ranked based on surrogate values. The cutoff applied at the beginning of the campaign is used again to select the $Top \; K\%$. The binary treatment decisions generated \textit{at the end of the campaign} serve as the surrogate ground truth on ``who should have been treated?". 


Notice how the SGT algorithm bypasses the impossibility of having an individual in the treatment and the control group simultaneously. The key is to fall back to the control (treatment) score of a treated (not treated) individual. This can be seen as one form of inferring outcomes of counterfactual scenarios. The \textit{true} lift score of a treatment individual is approximated using the observed KPI minus the prediction from the control model. The control score effectively accounts for the parallel scenario of the individual being in the control group. The first term induces zero error, since it is the actual empirical observation, while the second term inherits the error from the uplift model (and the model is the best proxy available). The reasoning for the control follows from symmetry.


One assumption we are making here is that the uplift models are consistent in ranking at time $t_\text{start}$ and at time $t_\text{end}$. Due to non-stationarity, a feature drift may occur between the start and end times. Keep in mind that the feature values are refreshed at time $t_\text{end}$ to provide the model with a better state representation to predict treatment/control scores. Clearly, the quality of SGT depends on the specific application, the predictive power of features, and the quality of uplift models. Despite that, our framework is generic and only requires simple sorting/ranking operations. The algorithm can be used to generate surrogate ground truth for any campaign design.
This provides two main benefits; i) evaluating the model performance given the surrogate labels, and ii) more importantly, enhancing binary fairness metric evaluations.

\smallskip

Finally, on a practical note, to benefit from enhanced fairness evaluations, we would evaluate fairness metrics using  ground-truth label from the SGT at a time $t_{end}$ that is still \textit{before} launching any live user-facing experimentation. 

\section{Experiments}
\label{Experimental}

The main goal of our experiments is to demonstrate how to apply surrogate ground truth generation in the context of campaign design in a real-world scenario. In particular, we study two research questions:

\begin{enumerate}
    \item[Q1] \textbf{Quality of labels generated by SGT:} An empirical assessment of the quality of the surrogate ground truth. Specifically, we compare the campaign profit obtained by the uplift model, the SGT, and a hypothetical Oracle.
    
    \item[Q2] \textbf{Fairness Evaluation:} Conduct an enhanced binary fairness evaluation, enabled by surrogate ground truth labels, and compare it to baseline fairness metrics that does not require ground truth.
\end{enumerate}

In the remainder of this section, we describe our dataset (\textsection Section~\ref{dataset}), the uplift modeling approach (\textsection Section~\ref{exp_model}), the performance of SGT (\textsection Section~\ref{exp_sgt}), and the result of enhanced binary fairness evaluation (\textsection Section~\ref{exp_metrics}).

\subsection{Campaign Dataset}\label{dataset}

While uplift modeling is extensively used in practice, it is worth noting that acquiring publicly available data sets for benchmarking is notoriously hard. To evaluate the quality of labels found by SGT requires ground truth, which defeats the purpose of generating SGT, and its lack thereof was our primary motivation to devise this approach. Additionally, industrial settings limit sharing real-world campaign data publicly, especially when it reveals sensitive information on protected attributes. For our purposes, we need a dataset that can provide two unique features: i) \textit{some} notion of ground truth so that we can evaluate the quality of SGT, and more importantly, ii) provide protected attributes to evaluate fairness metrics. 

\smallskip
The Starbucks Marketing Campaigns~\cite{lee2019p1,lee2019p2} dataset fits our experimental purposes neatly. First, this publicly available dataset provides information on $3$ different protected attributes; namely, gender, age, and income for a population of $17,000$ users. Secondly, it consists of $10$ different campaigns across various channels spanning a time horizon of $20$ months. This allows us to analyze SGT under various scenarios. Moreover, the historical campaign setup guarantees that only one campaign was active in any given month, which helps control the confounding effect. Finally, and most importantly, each customer exhibits both treatment and control behavior in each month, which can be considered as a reasonable  form to generate \textit{ground truth} which we can then compare to the ground truth SGT finds. The dataset provides the following marketing campaign types, customer profiles, and transactions.

\smallskip 

\begin{table}[t]
\centering{
\renewcommand{\arraystretch}{1.3}
  \begin{tabular}{cccccc}
    \toprule
    	C & Type & Channels & Difficulty & Duration & Reward \\
    	\hline
    1 & buy1-1free & [E,M,S] & 10 & 7 & 10 \\
    2 & buy1-1free & [W, E, M, S] & 10 & 5 & 10 \\
    3 & informational & [W, E, M] & 0 & 4 & 0 \\
    4 & buy1-1free & [W, E, M] & 5 & 7 & 5 \\
    5 & discount & [W, E] & 20 & 10 & 5 \\
    6 & discount & [W, E, M, S] & 7 & 7 & 3 \\
    7 & discount & [W, E, M] & 10 & 10 & 2 \\
    8 & informational & [W, E, S] & 0 & 3 & 0 \\
    9 & buy1-1free & [W, E, M, S] & 5 & 5 & 5 \\
    10 & buy1-1free & [W, E, M] & 10 & 7 & 2 \\
  \bottomrule
\end{tabular}
  \caption{Starbucks marketing campaign dataset as an instance of the Campaign Design Problem (CDP). Campaign dataset includes the campaign type (C), broadcasting channels from the web (W), email (E), mobile (M) and social (S), the difficulty captured as the amount customer should spend to activate the offer, the duration as the number of days the offer lasts, and the reward/cost value.}\label{tab:campaigns-descr}
}
\end{table}

\noindent \textbf{Portfolio of Marketing Campaigns:} As summarized in Table \ref{tab:campaigns-descr}, there are $10$ different campaigns ranging in type from buy-one-get-one-free, informational, to discount offers. Each campaign provides its broadcast channel (web, email, mobile, or social), difficulty as the amount of money spent for activation, offer duration, the reward it offers to customers, and conversely, the cost for the company.

\noindent \textbf{Customer Profiles:} There are $17,000$ de-identified customers with arbitrary customer IDs. For each customer, we are given information on age, gender, income, and day|month|year they became a member.

\noindent \textbf{Transactions:} There are $300,000$ transaction events classified as i) offer received, ii) offer viewed, iii) offer completed, and iv) transaction and its amount. 

\smallskip

\noindent \textbf{Preprocessing:}  Standard techniques are applied to impute missing values with zeros and remove duplicates. The transaction data was aggregated at the monthly level. Notice in Table \ref{tab:campaigns-descr} that the duration of campaigns ranges between $3$ to $10$ days. Effectively, each month provides a view into customers' promotional (treatment) and non-promotional (control) activity. The overall activity can be deemed profitable from the company's perspective, depending on the cost of the campaign. We normalized the profit generated for promotional and non-promotional settings to $30$ days as the promotional period is shorter in this setup ($3$-$7$ days in a month is promotional while the rest of the month is non-promotional). To represent customer context, time-based features are generated from transaction records. Precisely, cumulative sums and moving averages for several features up to the current month and $1$-month lag for each campaign were calculated. Features describing historical engagement with previous marketing campaigns were also included in the context. This led to $230$ dimensional feature vector from past transaction behavior.

\smallskip 

Table~\ref{tab:datasample} shows a concrete example with $2$ customers. Customer $1$ in month $3$ did not respond to campaign $2$ but purchased outside of the promotional window. Notice how the same customer exhibit both treatment and control behaviour within the same month. Customer $2$ in month $5$ spent money in the promotional time-window for campaign $4$ and also purchased outside of the promotional window. If an Oracle provided us with this view, we would not offer the promotion to the first customer since it was not profitable. This the type of oracle (ground truth) that we would like to revive using surrogates.  

\begin{table}
\renewcommand{\arraystretch}{1.3}
\centering{
  \begin{tabular}{ccccc}
    \hline
    	Customer & Month & Campaign Id & Purchase & Profitable \\
    	\hline
    1 & 3 & 2 & 0 & 0 \\
    1 & 3 & None & 17.15 & 1 \\
    2 & 5 & 4 & 31.78 & 1 \\
    2 & 5 & None & 19.92 & 1 \\
  \hline
\end{tabular}
  \caption{Example campaign data. Each customer in a campaign is represented by two rows each month: one for the promotional behavior and one for the non-promotional behavior measured by the purchase amount.}\label{tab:datasample}
}
 
\end{table}

\noindent \textbf{Train/Test Split}: We consider a time-based train-test split where earlier months of an active campaign serve as a training set, and later months serve as validation and test sets. We did not perform a user-based split since each user was subject to each campaign at maximum once. Thus, when splitting by time and campaign, we were guaranteed to not have the same user in both train and test sets. On average, this gives us \textasciitilde{}$7,500$ observations in training set over \textasciitilde{}$4$ months of data and roughly \textasciitilde{}$2,500$ observations in both the validation set and the test set over $1$ month of data each. The number of observations was similar across all marketing campaigns. 

\smallskip

\noindent \textbf{Model Building}: We consider a classification approach aimed at predicting customers' profitable status. Of note, our approach is also extensible to regression. The model outcome is converted to a binary yes/no decision about each individual's intervention as a standard setup in uplift modeling. Thus the underlying algorithm can be either estimating a propensity in classification settings or a continuous score in regression settings. 

\smallskip

The profitable label was highly imbalanced (ranging from $2$\% to $25$\% incidence rate in both treatment and control groups, with some variability across the $10$ marketing campaigns). Therefore, we utilize Synthetic Minority Oversampling Technique (SMOTE)~\cite{chawla2002smote} to up-sample the minority class to a $50/50$ ratio.

\smallskip

\begin{table}[t]
\centering{
\renewcommand{\arraystretch}{1.3}
  \begin{tabular}{cccccc}
    \hline
    	C & Active Months & Num. of Customers \\
    	\hline
    1 & [5, 11, 13, 16, 17, 19] & 5479 \\
    2 & [5, 11, 13, 16, 17, 19] & 5549 \\
    3 & [5, 11, 13, 16, 17, 19] & 5453 \\
    4 & [5, 11, 13, 16, 19] & 5457 \\
    5 & [5, 11, 13, 16, 19] & 5495 \\
    6 & [5, 11, 13, 16, 19] & 5440 \\
    7 & [5, 11, 13, 16, 19] & 5515 \\
    8 & [5, 11, 13, 16, 17, 19] & 5521 \\
    9 & [5, 11, 13, 16, 19] & 5410 \\
    10 & [5, 11, 13, 16, 19] & 5441 \\
  \bottomrule
\end{tabular}
  \caption{Campaign activity over time and different cohorts.} \label{tab:campaigns}
}
\end{table}

Due to the sparsity of time-based features, we use principal component analysis (PCA) to generate compact representations. On average, \textasciitilde{}$50$ components explain $90$\%+ of the variance observed, which is a considerable reduction from $230$ features.

\smallskip

Given the Treatment and Control groups, we proceed with training a two-model uplift approach. Similarly, other uplift modeling methods, such as the dummy-approach or the quadrant-approach, can be considered as defined in Section~\ref{uplift}. For each campaign, we train two XGBoost models with early stopping condition and hyperparameters as defined in ~\cite{lee2019p1,lee2019p2}. Let us stress again that our goal is not to build the most powerful uplift model for this dataset but to demonstrate how to utilize surrogate ground truth generation in practice.

\subsection{Uplift Modeling}\label{exp_model}

We study each campaign separately, decide the treatment and control groups, build the corresponding uplift models, and evaluate our SGT generation for each campaign. This gives a comprehensive evaluation of different scenarios. 

\noindent \textbf{Treatment \& Control Groups}: For each campaign, we consider the months that the campaigns were active throughout the $20$-month campaign horizon as given in Table~\ref{tab:campaigns}. We identify the promotional and non-promotional behavior of each customer during these months. The former yields the Treatment group and the latter yields the Control group.

\begin{table}[t]
\centering{
\renewcommand{\arraystretch}{1.3}
\begin{tabular}{cccccccc}
\hline
C & No Offer           & Full Offer & Oracle  & Uplift & SGT & Imp \\ 
\hline
1       & \$7,581              & \$523                            & \$7,942  & \$6,847                   & \$7,778  & \textbf{85\%} \\
2       & \$7,536              & \$5,004                          & \$12,020 & \$7,273                   & \$8,323  & 22\%                         \\
3       & \$7,796              & \$2,103                          & \$9,526  & \$7,415                   & \$8,261  & 40\%                         \\
4       & \$10,259             & \$1,020                          & \$11,254 & \$9,693                   & \$10,541 & 54\%                         \\
5       & \$8,738              & \$2,478                          & \$10,924 & \$8,257                   & \$9,097  & 32\%                         \\
6       & \$8,065              & \$5,511                          & \$13,345 & \$7,443                   & \$9,141  & 29\%                         \\
7       & \$8,927              & \$1,701                          & \$10,375 & \$8,297                   & \$9,401  & 53\%                         \\
8       & \$8,662              & \$2,745                          & \$11,126 & \$8,697                   & \$9,077  & \textbf{16\%}                         \\
9       & \$9,488              & \$3,112                          & \$12,262 & \$6,709                   & \$10,089 & 61\%                         \\
10      & \$9,096              & \$3,360                          & \$11,795 & \$7,925                   & \$9,943  & 52\%                         \\ \hline
& & & & & Avg &\textbf{ 44}\%  \\
\end{tabular}
\caption{Performance of different strategies @ Top-Decile.}\label{tab:performance}
}
\end{table}

\subsection{Performance of SGT [Q1]}\label{exp_sgt}

Given the uplift models, to answer Q1, we evaluate the performance of SGT for each of the 10 marketing campaign scenarios. Within each scenario, we experiment with different budget thresholds $B$ $\in \{5\%, 10\%, 15\%, 20\%\}$ and compare the following strategies applied to the test set, as summarized in Table
\ref{tab:performance}. 

\begin{itemize}
    \item \textbf{No Offer}: This strategy promotes no offers. Hence there is no promotion cost.
    \item \textbf{Full Offer}: This strategy promotes an offer for everyone in the test period, hence incurs cost for the entire population. 
    \item \textbf{Oracle}: Virtual best strategy that selects the most profitable option for everyone. In practice, this cannot be achieved. It gives us an upper bound on the expected performance to answer Q1.
    \item \textbf{Uplift}: The performance of the predictions made by the uplift model. In practice, this is the marketing campaign we would use. This is the user-impacting strategy we we want evaluate for potential bias. 
     \item \textbf{SGT}: Surrogate ground truth generated by Algorithm~\ref{algo}. We can evaluate SGT's performance by comparing it to Oracle (Q1), and we can evaluate the fairness of the uplift strategy by treating the SGT (or the Oracle) as the truth labels (Q2). 
\end{itemize}

Table \ref{tab:performance} presents the total profit generated by different strategies for the top decile (i.e., $B$ = $10$\%). Other budget scenarios exhibit similar results, hence omitted. The $Imp$ column shows the percentage of optimality gap closed by the SGT in the performance range between the model and the Oracle. The average improvement achieved by SGT across all campaign is given in Table~\ref{average-sgt}. When averaged over ten campaign scenarios, SGT closes $44\%$ of the gap toward the hypothetical best campaign for the top decile. 

As shown in Table~\ref{tab:performance}, promoting an offer for every individual in the test set is too costly. Consider campaign $C=1$: if no individual is offered a promotion, \$$7,581$ is generated in profit in the held-out test set. If every individual in the same set is offered a promotion, only \$$523$ in profit is generated leading to a profit reduction of -\$$7,058$. 
This is exactly the motivation behind using an uplift model to find truly responsive customers and generate profit greater than the "No Offer" baseline.  When the model's predictions are used to select the customers for the treatment, the profit increases steadily.

The uplift model in Table \ref{tab:performance} selects a $10$\% subset of population to send promotions. In campaign $C=1$, this leads to generating a profit of \$$6,847$. Such an uplift model would not be good enough in practice. In fact, it performs roughly on par with selecting $10$\% of population uniformly at random, assuming such selection would result in \$$7,058/10=\$705$ profit reduction. Overall, the Uplift Models in all $10$ campaigns except $C=8$ do not beat the "No Offer" baseline and would require further tuning to get closer to the best possible profit from that set, denoted by Oracle (which does have higher profit than the baseline "No Offer" strategy in all campaigns). It is important to note here that, in practice, proprietary datasets of companies have much richer  feature sets and hence likely to generate better signals than open-source datasets. 

\begin{table}[t]
\renewcommand{\arraystretch}{1.3}
\centering{
\begin{tabular}{cccc} \hline
 Budget & Max  & Min  & Mean \\ \hline
Top-20 & 94\% & 42\% & 65\% \\ 
Top-15 & 88\% & 24\% & 54\% \\ 
Top-10 & 85\% & 16\% & 44\% \\ 
Top-5  & 71\% & 4\%  & 26\% \\ \hline
\end{tabular}
\caption{The percentage of optimality gap closed by SGT over the uplift model when averaged over ten campaigns with varying campaign budget.}\label{average-sgt}
}
\end{table}

 While our goal was not to build the best model for this dataset, these results confirm a reasonable model performance: we beat the ``No Offer" baseline in one campaign. This dataset has a unique advantage to provide \textit{both} the promotional behavior and the non-promotional behavior of individuals (e.g., refer to Table~\ref{tab:datasample} again). The Oracle exploits this information leakage to select the best outcome for every decision. This gives an upper bound to assess the performance of SGT. The higher the percentage of the gap closed by SGT, the closer the SGT is to the hypothetical ground truth. For the top decile, Campaign $1$ emerges as the best case where SGT is able to improve over the model substantially and close $85\%$ of the gap. The worst-case occurs in Campaign $8$ where SGT closes only $16\%$ of the gap. The overall best and worst results were $94\%$ and $4\%$ across all Top-$K$ as shown in Table~\ref{average-sgt}. 
Notice the improving trend in SGT's performance as the budget constraint is relaxed from Top-5 to Top-20. The uplift model performs better in higher deciles and the optimality gap becomes loose with worsening model performance beyond the top semi-decile. SGT captures this effect and serves as proxy. When all budget constraints and all campaign scenarios are considered, the SGT closes \textasciitilde{}50\% of the performance gap on average. Keep in mind that, in reality, we would not have access to the Oracle.

Overall, SGT stands out as an attractive mechanism for generating a proxy for truth labels closing half of the Oracle gap on average. Once a campaign is over, it effectively answers the question of \textit{``who should have received the intervention?''}. This result on its own is a contribution of our paper for any CDP application.

\begin{table}[t]
\centering{
\begin{tabular}{c|cl|cl|cl|}
\cline{2-7}
\multicolumn{1}{l|}{}          & \multicolumn{2}{c|}{Age}           & \multicolumn{2}{c|}{Gender}        & \multicolumn{2}{c|}{Income}        \\
\multicolumn{1}{l|}{}          & \multicolumn{1}{l}{Base} & SGT    & \multicolumn{1}{l}{Base} & SGT    & \multicolumn{1}{l}{Base} & SGT    \\ \hline
\multicolumn{1}{|c|}{SP}       & \multicolumn{1}{l}{0.021} & 0.021  & \multicolumn{1}{l}{0.066} & 0.066  & \multicolumn{1}{l}{0.098} & 0.098  \\
\multicolumn{1}{|c|}{DI}       & \multicolumn{1}{l}{\color{red}{1.228}} & \color{red}{1.228}  & \multicolumn{1}{l}{\color{red}{1.859}} & \color{red}{1.859}  & \multicolumn{1}{l}{\color{red}{2.708}} & \color{red}{2.708}  \\
\multicolumn{1}{|c|}{AO}       & -                         & -0.003 & -                         & 0.04   & -                         & 0.052  \\
\multicolumn{1}{|c|}{EO}       & -                         & -0.026 & -                         & 0.008  & -                         & -0.022 \\
\multicolumn{1}{|c|}{FNR diff} & -                      d   & 0.026  & -                         & -0.008 & -                         & 0.022  \\
\multicolumn{1}{|c|}{PE}       & -                         & 0.021  & -                         & 0.073  & -                         & 0.127  \\ \hline
\end{tabular}
\caption{Fairness evaluation of the best performing SGT @ Top-Decile for Campaign $1$.}\label{fairness-best}}
\vspace{-0.4cm}
\end{table}

\vspace{-0.6cm}
\subsection{Enhanced Binary Fairness Evaluation [Q2]}\label{exp_metrics}
\vspace{-0.2cm}

Given the labels generated by SGT, we now turn to Q2 and binary fairness evaluation. We can evaluate the fairness of any campaign from Table~\ref{tab:performance} and let us consider the two extremes, the best and the worst campaigns in terms of SGT gap improvement, namely, Campaign $1$ and Campaign $8$. 

We evaluate the fairness with respect to protected attributes age, gender, and income. The interpretation of these metrics for age, gender, and income are as follows: an evaluation above the ideal value means a higher effect on older individuals, on females, and on individuals with high-income, respectively.

The results are presented in Table~\ref{fairness-best} and Table~\ref{fairness-worst}. The \textit{Base} evaluation includes only the metrics that do not depend on ground truth, whereas \textit{SGT} enables a more holistic view unlocking other metrics. Based on the ideal values and ranges from Table~\ref{metrics-summary}, potentially problematic values are shown in red color.

\begin{table}
\centering{
\begin{tabular}{c|cc|cc|cc|}
\cline{2-7}
\multicolumn{1}{l|}{}          & \multicolumn{2}{c|}{Age}                             & \multicolumn{2}{c|}{Gender}                          & \multicolumn{2}{c|}{Income}                          \\
\multicolumn{1}{l|}{}          & \multicolumn{1}{l}{Base} & \multicolumn{1}{l|}{SGT} & \multicolumn{1}{l}{Base} & \multicolumn{1}{l|}{SGT} & \multicolumn{1}{l}{Base} & \multicolumn{1}{l|}{SGT} \\ \hline
\multicolumn{1}{|c|}{SP}       & -0.006                    & -0.006                   & -0.027                    & -0.027                   & -0.04                     & -0.04                    \\
\multicolumn{1}{|c|}{DI}       & 0.942                     & 0.942                    & \color{red}{0.752}                     & \color{red}{0.752}                    & \color{red}{0.661}                     & \color{red}{0.661}                    \\
\multicolumn{1}{|c|}{AO}       & -                         & -0.014                   & -                         & -0.011                   & -                         & -0.04                    \\
\multicolumn{1}{|c|}{EO}       & -                         & -0.02                    & -                         & 0.013                    & -                         & -0.035                   \\
\multicolumn{1}{|c|}{FNR diff} & -                         & 0.02                     & -                         & -0.013                   & -                         & 0.035                    \\
\multicolumn{1}{|c|}{PE}       & -                         & -0.007                   & -                         & -0.036                   & -                         & -0.045                   \\ \hline
\end{tabular}
\vspace{-0.25cm}
\caption{Fairness evaluation of the worst performing SGT @ Top-Decile for Campaign $8$. }\label{fairness-worst}
}
\vspace{-0.5cm}
\end{table}

As shown in the Base column, without access to surrogate truth labels, we have a limited view on fairness evaluation: only Statistical Parity (SP) and Disparate Impact (DI) are available for fairness.

Based on this result, in Campaign $1$, we observe that the uplift model preferentially selects older females with higher income as truly responsive. In Campaign $8$, the uplift model preferentially selects males earning below-average income. These observations  raise the need for further evaluation, and SGT provides exactly that. We find that none of the other metrics, such as AO, EO, FNR diff, and PE, are outside the pre-described ideal ranges based on the ``80/20'' rule-of-thumb described in Section~\ref{metrics}.


We also examined the feature importance of variables used in the uplift models. We expect that if a protected attribute (age, gender, income) is among the critical features that models pick up, then evaluating potential bias becomes even more critical. We confirmed that income and age were among the most important features, closely followed by gender for both the treatment and control models. This further motivates the need for enhanced fairness evaluations. 


Comparing the per-group predictions via Disparate Impact (DI) we found values higher than ideal in the Table \ref{fairness-best} and Table \ref{fairness-worst}. Keep in mind that DI the a ratio of the number of positive predictions in membership groups, as defined in Equation~\ref{metric_di}. As such, it is sensitive to the number of members in each of the two groups. The metric can be skewed if the minority group has significantly fewer members, as is the case in this dataset. The enhanced fairness evaluation enabled by SGT sheds light on other metrics that turn out to be within ideal ranges. This boosts the confidence in the decisions of the uplift model. 

\section{Related Work}\label{related}

The key component of our method is to leverage the trained uplift model to get the control (treatment) score of a treated (control) individual in the counterfactual scenario. Thus, our method can be seen as a special case of doubly robust estimation (DRE) which is commonly used in  estimating counterfactuals for causal inference~\cite{489d9f8419ce4f059ffb84645531d34c}. The main difference is that, in DRE, new models need to be trained and scored once the campaign is over~\cite{dudik2011doubly}. This is not only an additional complexity for practitioners but also introduces new sources of bias and error. Compared to DRE, when generating the counterfactuals in the SGT algorithm, one of the terms induces zero bias, since it is the actual empirical observation, while the other term inherits the training bias from the uplift model (and the trained model is the best proxy available). When the goal is enhanced fairness evaluation, SGT can be viewed as simplification technique to allow practitioners conduct fairness evaluation without building new models. This is especially important for sparse data applications with limited observations where further splitting of campaign data into train/test sets could be detrimental for modeling. 

There exists related work that also study fairness evaluation in the context of treatment effect estimation and personalized decision-making. Similar to our work, ~\cite{nathan} considers the fundamental problem of causal inference where we cannot observe potential outcomes on sub-populations that remain un-treated. Unlike ~\cite{nathan}, SGT makes no assumption of monotone treatment response which is the cornerstone of their discussion. Briefly, the monotone 
 treatment assumption indicates that anti-responders, called do-not-disturbs in this paper, do not exist. This leads to an assumption that the treatment is either  not effective (per the example in ~\cite{nathan}; whether an individual receives a job offer, independent of receiving job training) or treatment benefits the individual (individual receives a job offer if and only if they receive the training) but it never harms the individual. We do not make such assumptions and maintain the four-quadrant outcomes matrix conditional on intervention as depicted in Figure~\ref{fig:one}. Hence, we are able to evaluate fairness of campaigns under the best possible selection to optimize the outcome of the overall campaign from the lens of the intervening agent (e.g., the company sending marketing offers which built its targeting model), while determining the fairness for each individual being treated. Additionally, the methods proposed in ~\cite{nathan}  can only be evaluated in randomized control trial settings while SGT works on evaluating campaign design problems where individuals were selected for treatment/control with a non-random model.

A different approach studied in~\cite{razieh}  focuses on adjusting the optimization process in learning algorithms such that there are fairness guarantees. That work is focused on learning fair policies in the context of reinforcement learning, Q-learning, value search and G-estimation, while our focus remains solely in the context of uplift modeling. Moreover, our approach to utilize SGT is significantly more straightforward since it does not require data science practitioners to implement their own custom loss functions with fairness-aware constraint optimization terms. In other words, our method works out of the box by the re-scoring operation using existing treatment and control models in the campaign. 

In a related direction for fairness guarantees, ~\cite{blossom} discusses a custom fairness-constrained offline bandit algorithm, called Robinhood, which guarantees a confidence level threshold of fairness which is defined by the user in a multi-arm bandit setup. The paper further posits that the design of the Robinhood algorithm that satisfies each of the three desired conditions is difficult. Our main difference to~\cite{blossom} is that we are not altering the decision-making algorithm, we are simply re-scoring control and treatment candidates with existing models akin to doubly robust estimation~\cite{dudik2011doubly}. Moreover, ~\cite{blossom} considers the multi-arm bandit settings, a form of reinforcement learning, not the uplift model settings where all decisions are made at the time of starting the campaign versus sequentially. 


A systematic discussion of existing uplift modeling methods can be found in~\cite{zhang2021unified}. Let us note again that, our SGT method to enable enhanced fairness evaluation in uplift models is closely  related to treatment effect estimation. For instance, when compared to X-Learner~\cite{zhang2021unified}, our re-scoring operation is similar the CATE estimation inputs, except we do not require building four different models but only depend on two estimators; one for treatment and one for control. Our
 Our unique contribution is the re-use of the already available treatment and control models (required by the uplift campaign in the first place) at the time of evaluation  to create proxy for ground truth labels and counterfactual estimates. The important realization is that, by design, we remain agnostic to the particular choice of the base learners. LEt us also point out that this can also be a limitatio as our core assumptionis that the intervention model is a good proxy of the behavior of the individuals in the dataset. Hence, SGT might not be particularly useful in scenarios where we do not have a dataset
 with enough signal to build a predictive predictive to achieve positive results. Future robustness studies might consider to theoretically quantify this dependency and potentially provide fairness guarantees based on campaign model performance.



Fairness has been studied within marketing and advertising applications. For instance,~\cite{sweeney2013} considers the case for racial bias in targeted online advertising while~\cite{hbsads2019} reports on bias in dynamic pricing models for personalized marketing. More recently,~\cite{wan2019addressing} examined bias in product recommendations. However, the evaluation of algorithmic fairness for \textit{uplift modeling} remains unexplored. While explainability of uplift models used in marketing has been studied (see e.g.,~\cite{xai2020}), evaluating fairness metrics remains limited, precisely due to the lack of ground truth. Our work is an attempt to close this gap in the literature.

\section{Conclusion}
\vspace{-0.2cm}
We presented a generic approach to evaluate fairness of uplift models where labels are not readily available by generating a proxy we call surrogate ground truth (SGT). Uplift models are commonly employed in medicine, marketing and advertising campaigns where fairness evaluation is critical since the decisions made impact people's experiences. Our SGT algorithm is generic to accommodate different ML models (here, we used XGBoost), and conceptually, only depends on a re-scoring operation once the campaign is over. 

We applied our method in a real-world scenario from Starbucks Promotional Offers covering ten different campaigns across four different channels. The generated surrogate labels served as an effective proxy for ground truth closing almost half of the optimality gap toward a hypothetical Oracle. Most importantly, the surrogate ground truth enabled an extensive fairness evaluation beyond the standard baseline approach.


For future work, from the uplift modeling perspective, the robustness and sensitivity of generating SGT should be studied further. Directions include studying the effect of sample size and positive label ratio on SGT quality. Furthermore, predictions are susceptible to noise and point estimates might not be reliable~\cite{pachamova2020}. 
From the fairness perspective, group-based fairness is weak against composition with not enough samples in the minority group. Beyond group-based fairness, further research is needed to examine whether uplift models are desirable from the perspective of every individual. Finally, extending SGT to multi-class classification and regression is also of great interest. 

\newpage
\section*{Conflict of Interest Statement}

The authors have no competing interests or conflicts to declare that are relevant to the content of this article.

\section*{Author Contributions}
All authors contributed equally to this study and the write-up of the manuscript. All authors read and approved the final manuscript.

\section*{Funding}
No funding was received other than the support of the authors' employer. 



\section*{Data Availability Statement}
The dataset used in our experiments are public benchmarks that are available from the references therein. The fairness evaluation library, Jurity, is open-source on GitHub.


\begin{itemize}
    \item Starbucks Campaign Dataset - I:
    \href{https://towardsdatascience.com/implementing-a-profitable-promotional-strategy-for-starbucks-with-machine-learning-part-1-2f25ec9ae00c}
    {https://towardsdatascience.com/implementing-a-profitable-promotional-strategy-for-starbucks-with-machine-learning-part-1-2f25ec9ae00c}
    
     \item Starbucks Campaign Dataset - II:
     \href{https://towardsdatascience.com/implementing-a-profitable-promotional-strategy-for-starbucks-with-machine-learning-part-2-8dd82b21577c}
    {https://towardsdatascience.com/implementing-a-profitable-promotional-strategy-for-starbucks-with-machine-learning-part-2-8dd82b21577c}
     
     \item Jurity: Fairness \& Evaluation Library
     \\\url{https://github.com/fidelity/jurity}
     
\end{itemize}


\bibliographystyle{plain}
\bibliography{references}

\end{document}